\documentclass[english]{revtex4}
\usepackage{times}
\usepackage[T1]{fontenc}
\usepackage[latin1]{inputenc}
\usepackage{geometry}
\usepackage{graphicx}
\usepackage{amssymb}

\makeatletter

\usepackage{babel}
\makeatother
\begin{document}

\title{Mapping Luminosity-Redshift Relationship to LTB Cosmology}

\author{Daniel J. H. Chung and Antonio Enea Romano}

\affiliation{Department of Physics, University of Wisconsin, Madison, WI 53706,
USA}

\begin{abstract}
We derive a direct general map from the luminosity distance $D_{L}(z)$
to the inhomogeneous matter distribution $M(r)$ in the
Lemaitre-Tolman-Bondi (LTB) cosmology and compute several
examples. One of our examples explicitly demonstrates that it is
possible to tune the LTB cosmological solution to approximately
reproduce the luminosity distance curve of a flat FRW universe with a
cosmological constant. We also discuss how smooth matter distributions
can evolve into naked singularities due to shell crossing when the
inhomogeneous ``curvature'' $E(r)$ is a function which changes sign.
\end{abstract}
\maketitle

\section{Introduction}

Standard Friedmann-Robertson-Walker (FRW) cosmology is
characterized by the following features: homogeneity and isotropy when
averaged on $\mathcal{O}(100)$ Mpc length scales, negligible spatial
curvature, and a stress tensor which today has a chemical composition
of approximately 70\% dark energy and 30\% pressureless dust. The standard
inflationary embedding of FRW cosmology currently provides a successful picture of the universe.  However, since the invocation of
dark energy leads to a new coincidence problem, which is that dark
energy dominance roughly coincides with the epoch of nonlinear
structure formation, and since we still do not understand 
the cosmological constant problem, there has been a renewed interest
in exploring alternative inhomogeneous cosmological models (which are
not perturbatively related to FRW cosmologies) to see whether they
might offer a competitively plausible picture of the universe
\cite{Alnes:2005rw,Celerier:1999hp,Vanderveld:2006rb,Chuang:2005yi,Moffat:2005yx,Moffat:2005ii,Nambu:2005zn,Ishibashi:2005sj,Apostolopoulos:2006eg,Kai:2006ws,Biswas:2006ub}.

Supernova luminosity distance measurements as a function of redshift
offer compelling evidence for dark energy when interpreted in the
context of FRW cosmologies
\cite{Perlmutter:1998np,Riess:1998cb}.  To interpret such data in the
context of inhomogeneous cosmologies, it is often not particularly useful to average the underlying inhomogeneous variables to obtain a
forced interpretation in terms of FRW cosmology.  This is because
generically there is no preferred spatial slicing to compute averages
and no meaningful map between the physical observables and the time
derivatives of the spatially averaged variables. In some sense, the
observables contain more information than that which could be carried
by averaged variables, and hence determining the appropriate smearing
map on the observables which matches the information that could be carried
by averaged variables requires a knowledge of the underlying
inhomogeneities in the absence of special symmetries. Therefore, to
characterize inhomogeneous cosmologies, it is useful to compute observables
such as the luminosity distance function directly in terms of
variables describing the underlying inhomogeneous geometry.

The Lemaitre-Tolman-Bondi (LTB) solution corresponds to a spherically
symmetric exact solution to the Einstein equations with pressureless
ideal fluid. Extensive analyses have been carried out for this model
because it allows for investigations of inhomogeneities that cannot be
represented as perturbative deviations from FRW cosmologies. The LTB
solution is fixed by choosing three smooth functions $\{
E(r),M(r),R_{0}(r)\}$, which allow an infinite number of different
radial inhomogeneities. Most of the previous attempts to compute
a cosmologically plausible luminosity distance $D_{L}(z)$ as a function
of redshift $z$ consisted of finding a map from $\{
E(r),M(r),R_{0}(r)\}$ to $D_{L}(z)$ or a Taylor expansion of $D_L(z)$
about $z=0$
\cite{Celerier:1999hp,Alnes:2005rw,Sugiura:1999dp,Mustapha:1997xb}.

In this paper, we derive a map from $\{ E(r),D_{L}(z),R_{0}(r)\}$ to
$M(r)$. This has the advantage that the observed luminosity distance
function $D_{L}(z)$ more directly dictates the underlying cosmological
model, as opposed to having to guess the right $\{
E(r),M(r),R_{0}(r)\}$ to produce the desired $D_{L}(z)$. In
\cite{Vanderveld:2006rb}, a similar inverse problem (with a different
choice of variables) was considered which focused on the situation
with $E=0$, while in this paper, we will keep $E$ general. In using
this new map, we find the interesting fact that the luminosity
distance function is typically (depending on the choice of $E(r)$) an
effective probe of the LTB geometry only for $z\lesssim1$, since the
luminosity distance function has a universal behavior in the limit
that $\{ R_{0}=0,E\rightarrow0\}$.  More precisely, $D_{L}(z)$ is
fixed independently of $M(r)$ for $\{ R_{0}=0,E=0\}$, since that limit
corresponds to the $\Omega_M=1$ FRW universe.  A negative
consequence of this feature is that the differential equation map from
$\{ E(r),D_{L}(z),R_{0}(r)\}$ to $M(r)$ fails to be a numerically
accurate map beyond $z \sim 1$.  Nonetheless, we show how the
numerical solution can be patched to a semi-analytic solution beyond
$z=1$ to obtain a good fit (to within around $5\%$ for the redshifts of
interest) to obtain an LTB cosmology which reproduces the luminosity
distance function of an FRW cosmology with $\Omega_{\Lambda}=0.7$ and
$\Omega_{M}=0.3$.

Furthermore, we make a subsidiary observation about the class of LTB
solutions which can mimic the observed $D_{L}(z)$. We find that if the
radial inhomogeneity profile goes from $E(r)>0$ to $E(r)<0$ as $r$ increases
while $M'(r)$ is positive in that region, there is generically a
danger of forming naked singularities, which can be interpreted as due
to shell crossing.

The order of presentation will be as follows.  In the next section, we
review the conventional approach to obtaining the luminosity distance
as a function of redshift in the LTB cosmologies.  In
Sec.~\ref{sec:Inversion-method}, we construct a set of differential
equations (which we will refer to as the inversion method) which can
be used to map the luminosity distance function into a particular LTB
geometry.  Afterwards, we apply the method to several examples. In
Sec.~\ref{sec:Naked-Singularity}, we discuss how smooth geometries can
evolve into naked singularities when $E(r)$ switches signs.  We then
summarize and conclude.  For completeness, we present the FRW
luminosity distance with $\Omega_M+\Omega_\Lambda=1$ in Appendix A. In
Appendix B, we write the LTB solution explicitly in a form that is not
commonly found in the literature.

\section{Conventional Approach\label{sec:Conventional-Approach}}
The spherically symmetric Lemaitre-Tolman-Bondi (LTB) metric\begin{equation}
ds^{2}=dt^{2}-\frac{(R_{,r})^{2}}{1+2E(r)}dr^{2}-R^{2}d\Omega_{2}\end{equation}
satisfies the Einstein equation with\begin{equation}
T_{\,\,\nu}^{\mu}=\textrm{Diagonal}[\rho=\frac{M_{pl}^{2}}{4\pi}\frac{M'(r)}{R^{2}R_{,r}},-P=0,-P=0,-P=0]\label{eq:energydensity}\end{equation}
for differentiable functions $E(r)$ and $M(r)$ if $R(t,r)$ satisfies
the partial differential equation\begin{equation}
\left(\frac{\partial_{t}R}{R}
\right)^{2}=\frac{2E(r)}{R^{2}}+\frac{2M(r)}{R^{3}}.\label{eq:einstein00orig}\end{equation}
The function $E(r)$ can be thought of as a generalized version of the
spatial curvature parameter ($E(r) \propto -kr^{2}$ in FRW), while $M(r)$ can
be thought of as a generalized version of mass (for matter domination
in FRW $M(r) \propto \rho_{i}a_{i}^{3}r^{3}/M_{pl}^2$, where $\rho_{i}$ is an initial
energy density, $a_i$ is an initial scale factor, and $r$ is the
radial coordinate).
The function $R$ in the FRW limit takes the form $ra(t).$ The boundary
condition for Eq.  (\ref{eq:einstein00orig}) is provided by the radial
function $R_{0}(r)\equiv R(t_{0},r)$, where $t_{0}$ is the time at
which the boundary condition is set (note that this time is
generically different from today).

The luminosity distance in LTB model is approximately given \cite{Celerier:1999hp,Kristian:1965sz}
by\begin{equation}
D_{L}(z)=(1+z)^{2}R(t(z),r(z))\label{eq:lumindist1}\end{equation}
\begin{equation}
\frac{dr}{dz}=\frac{\sqrt{1+2E(r(z))}}{(1+z)\partial_{t}\partial_{r}R(t(z),r(z))}\label{eq:lumindist2}\end{equation}
\begin{equation}
\frac{dt}{dz}=\frac{-|\partial_{r}R(t(z),r(z))|}{(1+z)\partial_{t}\partial_{r}R(t(z),r(z))},\label{eq:lumindist3}\end{equation}
where $t(z)$ and $r(z)$ physically represent the geodesic of the
photon coming to us (located at $r=0$) starting from the radial
distance of our horizon. Any photon we observe that starts from a
closer radial distance will have a redshift which is the same as that
experienced by the horizon photon when the ratio of the frequencies
measured from any closer radial position is accounted for since 
redshift is independent of the frequency. Given an arbitrary choice of
$M(r)$ and $E(r)$, Eqs.~(\ref{eq:lumindist2}) and
(\ref{eq:lumindist3}) are conventionally solved to obtain the
luminosity distance function through
Eq.~(\ref{eq:lumindist1}). Angular diameter distance can be obtained
from the luminosity distance function by dividing by $(1+z)^{2}$.

One can simplify Eqs. (\ref{eq:lumindist2}) and (\ref{eq:lumindist3})
in the regime in which $\partial_{t}R$ ({\it i.e.}, the {}``local
expansion'' rate) maintains the same sign by rewriting
Eq. (\ref{eq:einstein00orig}) as\begin{equation}
\partial_{t}R=s\sqrt{2E(r)+\frac{2M(r)}{R(t,r)}},\label{eq:einstein00}\end{equation}
where $s\equiv\pm1$ specifies whether there is local expansion or
contraction. The solution to this differential equation requires a
specification of a function of $r$ at the initial time hypersurface.
We will define that function to be $R_{0}(r)$: $R_{0}(r)\equiv
R(t_0,r)$ (recall $t_0$ is not necessarily today).  Hence, we
compute\begin{equation}
\partial_{t}\partial_{r}R(t(z),r(z))=s\frac{E'(r)+M'/R-M\partial_{r}R/R^{2}}{\sqrt{2E(r)+\frac{2M}{R(t,r)}}},\end{equation}
and rewrite Eqs. (\ref{eq:lumindist2}) and (\ref{eq:lumindist3})
as\begin{equation}
\frac{dr}{dz}=\frac{s\sqrt{1+2E(r(z))}\sqrt{2E(r)+\frac{2M}{R(t,r)}}}{(1+z)[E'(r)+M'/R-M\partial_{r}R/R^{2}]}\label{eq:lumindist2pr}\end{equation}
\begin{equation}
\frac{dt}{dz}=\frac{-s|\partial_{r}R(t(z),r(z))|\sqrt{2E(r)+\frac{2M}{R(t,r)}}}{(1+z)[E'(r)+M'/R-M\partial_{r}R/R^{2}]}.\label{eq:lumindist3pr}\end{equation}
These have the advantage that there are no second derivatives appearing
in the equations, but have the added assumption that the sign $s$
is a constant.

\section{\label{sec:Inversion-method}Inversion method}
In Sec. \ref{sec:Conventional-Approach}, we explained the conventional approach
of obtaining the luminosity distance function $D_{L}(z)$ for a given
$\{ M(r),E(r),R_{0}(r)\}$. In this section, we wish to stipulate $D_{L}(z)$ and solve for the class of $\{ M(r),E(r),R_{0}(r)\}$
that corresponds to this luminosity distance. In particular, we will
solve for $M(r)$ for a given $\{ E(r),D_{L}(z),R_{0}(r)\}$. This
inversion method has the advantage that the physical observable $D_{L}(z)$
can be mapped to the geometry of the underlying model directly without
having to guess $M(r)$. The physics is simply that if one knows a
single radial geodesic history of a photon which was emitted at an
event $(t_{1},r_{1})$ and observed at $(t_{2},r_{2}),$ one knows
the full spacetime geometry in the region $(t_{1}<t<t_{2},r_{1}<r<r_{2})$
of the LTB solution owing to its spherical symmetry.

The basic equations for this goal also stem from Eqs. (\ref{eq:lumindist2pr}),
(\ref{eq:lumindist3pr}), and the equation for luminosity distance,
Eq. (\ref{eq:lumindist1}):\begin{equation}
R(z)\equiv R(t(z),r(z))=\frac{D_{L}(z)}{(1+z)^{2}}.\end{equation}
 Since Eqs. (\ref{eq:lumindist2pr}) and (\ref{eq:lumindist3pr})
depend on $\partial_{r}R$, we would like to find an expression for
$\partial_{r}R$ as a function of $z$. To this end, we use the exact
solution, which is given in Eq. (\ref{eq:maineqeinst}):

\begin{equation}
-(t-t_{0})\sqrt{2}E(r)+\sqrt{R(t,r)}\sqrt{E(r)R(t,r)+M(r)}-Q(r)=\frac{M(r)}{\sqrt{E(r)}}\ln[\frac{\sqrt{R(t,r)}+\sqrt{\frac{M(r)}{E(r)}+R(t,r)}}{\sqrt{R_{0}(r)}+\sqrt{\frac{M(r)}{E(r)}+R_{0}(r)}}]\label{eq:maineq}\end{equation}
\begin{equation}
Q(r)\equiv\sqrt{R_{0}(r)}\sqrt{E(r)R_{0}(r)+M(r)}\end{equation}
(or Eq. (\ref{eq:E<0}) if $E<0$). Taking
$\partial_{r}$ of Eq. (\ref{eq:maineq}), we obtain a linear equation
for $\partial_{r}R$, 
which can be solved to find\begin{eqnarray}
\partial_{r}R & = & \frac{f[R]\sqrt{R_{0}}}{f[R_{0}]\sqrt{R}}R_{0}'(r)+\frac{E'(r)f[R]}{2E\sqrt{R}}\left(2\sqrt{2}(t-t_{i})+\frac{R_{0}^{3/2}}{f[R_{0}]}-\frac{R^{3/2}}{f[R]}-\frac{M_{1}}{E^{3/2}}\ln\frac{\sqrt{ER}+f[R]}{\sqrt{ER_{0}}+f[R_{0}]}\right.\nonumber \\
 &  & \left.+\frac{M_{1}^{2}}{E^{3/2}}\left[\frac{1}{M_{1}+ER_{0}+\sqrt{ER_{0}}f[R_{0}]}-\frac{1}{M_{1}+ER+\sqrt{ER}f[R]}\right]\right)\label{eq:partialr}\\
 &  & +\frac{M'(r)}{2E}\left(-1+\frac{f[R]\sqrt{R_{0}}}{f[R_{0}]\sqrt{R}}+\frac{f[R]}{\sqrt{ER}}\left[2\ln\frac{\sqrt{ER}+f[R]}{\sqrt{ER_{0}}+f[R_{0}]}+\frac{M_{1}}{f^{2}[R]+\sqrt{ER}f[R]}-\frac{M_{1}}{f^{2}[R_{0}]+\sqrt{ER_{0}}f[R_{0}]}\right]\right)\nonumber \end{eqnarray}
where\begin{equation}
f[X]\equiv\sqrt{M_{1}+E(r(z))X}\end{equation}
\begin{equation}
M_{1}(z)\equiv M(r(z)).\end{equation}
Furthermore, the function $M'(r)$ in Eqs. (\ref{eq:lumindist2pr})
and (\ref{eq:lumindist3pr}) can be replaced by\begin{equation}
M'(r)=\frac{dM_{1}}{dz}/\frac{dr}{dz}.\end{equation}
Since there are three unknown functions $\{ M_{1}(z),r(z),t(z)\}$,
and Eqs.~(\ref{eq:lumindist2pr}) and (\ref{eq:lumindist3pr}) (with
appropriate substitutions for $\partial_{r}R$ and $M(r)$) provide
only two independent equations, we require another independent equation.
This is provided by $dR/dz$ through the chain rule: 

\begin{eqnarray}
\frac{d}{dz}R & = & s\sqrt{2E+\frac{2M_{1}}{R}}\frac{dt}{dz}+\partial_{r}R\frac{dr}{dz}.\label{eq:dRdzchainrule}\end{eqnarray}
For a given set of $\{ E(r),D_{L}(z),R_{0}(r)\}$, the set of differential
equations Eq.~(\ref{eq:lumindist2pr}), Eq.~(\ref{eq:lumindist3pr}), and
Eq.~(\ref{eq:dRdzchainrule}) can be solved for $\{ t(z),r(z),M_{1}(z)\}$.
Finally, to obtain $M(r)$, we invert $r(z)$ to obtain \begin{equation}
M(r)=M_{1}(z(r)).\end{equation}
In practice, as we discuss below, the procedure we just described
is a bit more difficult because the differential equation can become
singular for certain choices of $\{ E(r),D_{L}(z),R_{0}(r)\}$. Also,
for numerical implementation, it is useful to write Eqs. (\ref{eq:lumindist2pr}),
(\ref{eq:lumindist3pr}), and (\ref{eq:dRdzchainrule}) in the form\begin{equation}
\left(\begin{array}{c}
\frac{dt}{dz}\\
\frac{dr}{dz}\\
\frac{dM_{1}}{dz}\end{array}\right)=\left(A\right)
\label{eq:numdiffeqs}
\end{equation}
where $\left(A\right)$ is a matrix that does not contain any derivative
terms. However, this matrix contains hundreds of terms
consisting of combinations of $\{ E,E',D_{L},\frac{d}{dz}D_{L},M_{1},R_{0}\}$,
and is not very illuminating in the general case. 

Regarding the initial conditions, note that because the
differential equation also generically has a $0/0$ division near
$r=0$, a numerical implementation must typically set the boundary
condition at a small but nonvanishing $r$.  For this purpose, it is
useful to linearize the system about $r=0$ to obtain intuition
about the boundary condition near the origin.  If we assume that
$R(t,0)=0$, $E(r=0)=0$, $\lim_{r \rightarrow 0} \frac{M}{R}=0$, and
$R(t,r)\approx r \partial_r R(t,0)$ near $r=0$, we find
\begin{equation}
t-t_i = -z \frac{d D_L}{dz}|_{z=0} + \mathcal{O}(z^2)
\label{eq:tlinear}
\end{equation}
where $t_i$ is the value of $t$ at $z=0$.  Unfortunately, similar
expressions for $M_1(z)$ and $r(z)$ near $z=0$ depend upon the
assumption of the scaling behavior of $M(r)$ and $E(r)$ near $r=0$,
and even the limiting expressions are algebraically complicated partly
because of the presence of logs.  For example, taking the ansatz
$E(r)\propto r^2$, $r=r'(0)z$, and $M_1(z)=M_c z^3$, we obtain 
from Eq.~(\ref{eq:numdiffeqs}) a nonlinear consistency equation near
$z=0$ for $\{r'(0),M_c\}$.  Hence, it is simpler to directly solve for the
initial conditions numerically using Eq.~(\ref{eq:tlinear}), the ansatz
$r=r'(0)z$, and Eq.~(\ref{eq:numdiffeqs}) in the limit $z \rightarrow
0$.

\section{Examples}
In this section, we will solve Eqs.~(\ref{eq:lumindist2pr}),
(\ref{eq:lumindist3pr}), and (\ref{eq:dRdzchainrule}) for a variety of
choices of $\{ E(r),D_{L}(z),R_{0}(r)\}$ to demonstrate the inversion
method described in Section \ref{sec:Inversion-method} in physically
relevant examples.

\subsection{Flat FRW example\label{sub:FRW-example}}

As a first example of mapping $\{ D_{L}(z),E(r)\}$ to $\{ M(r),R(t,r)\}$,
consider the matter dominated ($\Lambda=0$) flat Friedmann-Robertson-Walker
(FRW) universe\begin{equation}
\{ R_{0}(r)=0,E(r)=0\}\end{equation}
with $s=1$. The function $\partial_{r}R$ can be found from Eq. (\ref{eq:partialr})
as\begin{equation}
\partial_{r}R(t,r)=\frac{1}{3}\frac{M'(r)}{M(r)}R(t,r).
\label{eq:partialrRinFRWlimit} \end{equation}
The resulting differential equations (Eqs. (\ref{eq:lumindist2pr}),
(\ref{eq:lumindist3pr}), and (\ref{eq:dRdzchainrule})) are\begin{eqnarray}
\frac{dr}{dz} & = & \frac{\sqrt{\frac{2M}{R(t,r)}}}{(1+z)[\frac{1}{\frac{dr}{dz}}\frac{d}{dz}M/R-\frac{1}{3}M'/R]}\\
 & = & \frac{\sqrt{2M_{1}R}}{(1+z)[\frac{2}{3}\frac{d}{dz}M_{1}]}\frac{dr}{dz}\end{eqnarray}
\begin{eqnarray}
\frac{dt}{dz} & = & \frac{-R^{3/2}}{(1+z)\sqrt{2M_{1}}}\end{eqnarray}
\begin{eqnarray}
\frac{dR}{dz} & = &
 \frac{1}{3}\frac{\frac{d}{dz}M_{1}}{M_{1}}R+\sqrt{\frac{2M_{1}}{R}}\frac{dt}{dz}\nonumber
 \\ & = & [1-\sqrt{\frac{2M_{1}}{R}}]\frac{R}{3}\frac{d}{dz}\ln
 M_{1}\label{eq:dRdzcapital}.\end{eqnarray} Note that the $dr/dz$
 equation in this limit becomes independent of $dr/dz\neq0$ because of
 the absence of $E$. Hence, this equation can be written
 as\begin{equation} (1+z)[\frac{1}{3}\frac{d}{dz}\ln
 M_{1}]=\sqrt{\frac{R}{2M_{1}}}.\label{eq:drdzsimplified}\end{equation}
 Note that Eqs. (\ref{eq:dRdzcapital}) and (\ref{eq:drdzsimplified})
 give\begin{equation}
 R=c\frac{M_{1}^{1/3}}{1+z}\label{eq:Rzdetermined}\end{equation} where
 $c\equiv R(z=0)/M_{1}^{1/3}(z=0)$. Solving for $M_{1}$ and $R$, we
 find\begin{eqnarray*} M_{1} & = &
 2^{3/2}c^{3/2}\left(1-\sqrt{\frac{1}{1+z}}\right)^{3}\end{eqnarray*}
 and\begin{equation}
 R=\frac{\sqrt{2}c^{3/2}}{(1+z)}\left(1-\sqrt{\frac{1}{1+z}}\right).\label{eq:FRWluminosity}\end{equation}
 Remarkably, $M(r(z))=M_{1}(z)$ is fixed without specifying $r(z)$. Note also that if we identify
 $\sqrt{2}c^{3/2}=\frac{2}{H_{0}}$, this $R(z)$ corresponds to the
 function implied by the luminosity distance of an $\Omega_{M}=1$ FRW
 universe.  Indeed, inserting Eq.~(\ref{eq:partialrRinFRWlimit})
 into the expression for $T_{00}$ in Eq.~(\ref{eq:energydensity}), one finds that the stress tensor
 corresponds to a homogeneous FRW universe: $T_{00} \propto t^{-2}$
 independently of $r$.  Hence, this limit corresponds to the
 homogeneous matter density FRW limit.

If we choose \begin{equation}
r(z)\propto(1-\frac{1}{\sqrt{1+z}}),\end{equation} which corresponds
to the geodesic-redshift relationship in an FRW universe, we
obtain\begin{equation} M(r)=M_{1}(z(r))\propto r^{3},\end{equation}
which corresponds to the $M(r)$ leading to the familiar FRW solution. From
Eq. (\ref{eq:maineq}), we can solve for $R$ finding $R\propto ar.$
However, note that we can choose another $r(z)$ and obtain an infinite
class of different $M(r)$ functions corresponding to the same
luminosity distance function implied by Eq. (\ref{eq:FRWluminosity}).

We can turn this result around. As long as $M/R\gg E$, the luminosity
distance curve no longer accurately probes the geometry of the LTB
model since different geometries lead to approximately the same $R(z)=D_{L}(z)/(1+z)^{2}$.
In particular, this means the inversion method will necessarily be
unstable once the curvature term $E$ can be neglected. Schematically,
we will have\begin{equation}
\frac{dr}{dz}\sim\frac{ER}{M_{1}}F+\frac{\sqrt{2M_{1}R}}{(1+z)[\frac{2}{3}\frac{d}{dz}M_{1}]}\frac{dr}{dz},\end{equation}
 when $ER/M_{1}$ becomes small and $F\sim\mathcal{O}(\frac{dr}{dz})$.
Since $\frac{\sqrt{2M_{1}R}}{(1+z)[\frac{2}{3}\frac{d}{dz}M_{1}]}\sim1$
in the limit that $ER/M_{1}\rightarrow0$, we have\begin{equation}
\frac{dr}{dz}\sim\frac{ER/M_{1}}{1-\frac{\sqrt{2M_{1}R}}{(1+z)[\frac{2}{3}\frac{d}{dz}M_{1}]}}F\sim\frac{0}{0},\end{equation}
which creates an unstable differential equation for $\frac{dr}{dz}$
in this limit. 

\subsection{\label{sub:Void-Model}Void Model}
We will now reproduce the void model of \cite{Alnes:2005rw}
by applying the inversion method as a nontrivial check. 

First, let us review the solution of \cite{Alnes:2005rw}. Their ansatz
for $E(r)$ and $M(r)$ can be written as\begin{equation}
E(r)=\frac{1}{2}H_{\perp,0}^{2}r^{2}(\beta_{0}-\frac{\Delta\beta}{2}[1-\tanh\frac{r-r_{0}}{2\Delta r}])\label{eq:eansatz}\end{equation}
\begin{equation}
M(r)=\frac{1}{2}H_{\perp,0}^{2}r^{3}(\alpha_{0}-\frac{\Delta\alpha}{2}[1-\tanh\frac{r-r_{0}}{2\Delta r}])\label{eq:mansatz}\end{equation}
\begin{equation}
R(t_{rec},r)=a_{rec}r\label{eq:rinitcondagain},\end{equation}
with\begin{equation}
\{\alpha_{0}=1,\beta_{0}=0,\Delta\beta=-\Delta\alpha=-0.9,\Delta r=0.4r_{0},r_{0}\approx\frac{1}{5H_{0}},H_{\perp,0}\approx H_{0}\},\label{eq:params}\end{equation}
where $H_{0}\approx50\textrm{km/s/Mpc}$ and $a_{rec}\sim10^{-3}$
is the effective scale factor at recombination. Note that in their
solution, $E(r)>0$ never changes sign. Hence, the solution is always
in a {}``locally open'' universe (recall $E(r)\sim-kr^{2}$ in the FRW
limit). The resulting luminosity distance $D_{L}^{\textrm{void}}(z)$
fits the supernovae luminosity distance well \cite{Alnes:2005rw}.

\begin{figure}
\begin{center}\includegraphics{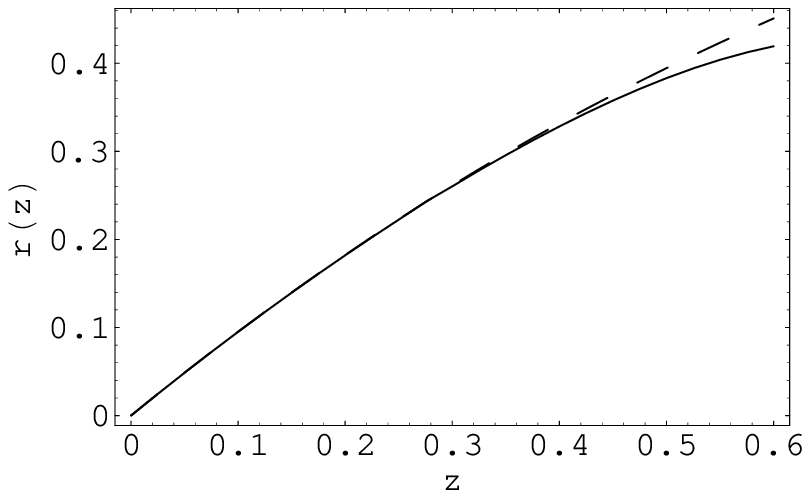}\includegraphics{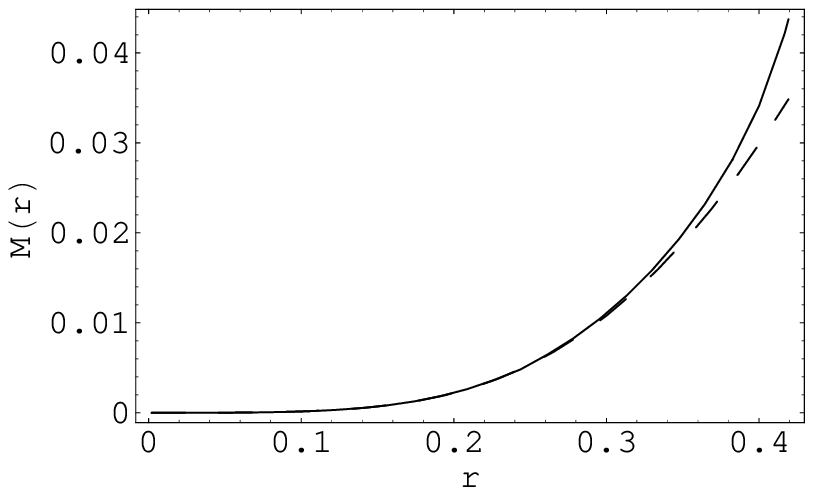}\end{center}

\caption{\label{rofz}In the graph on the left, the solid curve shows
the reconstructed $r(z)$ (in units of $1/H_0$) through the inversion
method and the dashed curve shows the $r(z)$ that would be obtained if
$M(r)$ of Eq.~(\ref{eq:mansatz}) were given as an input.  The
breakdown of the inversion method for this model around $z\approx0.6$
is apparent and is explained in the text. In the graph on the right,
the solid curve gives the $M(r)$ as determined through the numerical
evaluation of $M_{1}(z(r))$, and the dashed curve gives $M(r)$ as given by
Eq.~(\ref{eq:mansatz}).}
\end{figure}

To use the inversion method to derive this $M(r)$, we set $E(r)$
to Eq.~(\ref{eq:eansatz}), $R(z)=D_{L}^{\textrm{void}}(z)/(1+z)^{2}$,
and $R_{0}(r)$ equal to Eq.~(\ref{eq:rinitcondagain}) and solve
the differential equations Eqs.~(\ref{eq:lumindist2pr}), (\ref{eq:lumindist3pr}),
and (\ref{eq:dRdzchainrule}) subject to the boundary condition $\{
t(z_{i})\approx0.855\frac{1}{H_{0}},r(z_{i})\approx0,M_{1}(z_{i})\approx0\}$, 
where $z_{i}\approx0$ and $\lim_{r\rightarrow0}M_{1}(z(r))/r^{3}\approx0.084H_{0}^{2}$
was taken to match Eq.~(\ref{eq:params}) \footnote{The value of
  $t(z_i)$ is chosen for numerical convenience and is essentially arbitrary since it merely corresponds to a shift of the origin of time.}.
These boundary conditions correspond to ray tracing starting from
the {}``center of the universe'' where observations are assumed
to be made.

\begin{figure}
\begin{center}\includegraphics{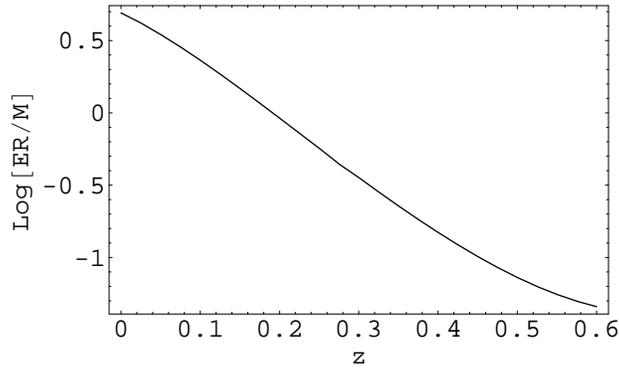}\end{center}

\caption{\label{cap:-ER/M-}$ER/M$ versus photon redshift $z$. For $z=0.6$, $ER/M\sim10^{-2}$ and the $dr/dz$
differential equation is unstable. Another way to view the instability is that
the luminosity distance function is not sensitive to the geometry
for $z\gtrsim0.6$ in this model. }
\end{figure}

The results of the $r(z)$ and $M(r)$ reconstruction are shown in
Fig.~\ref{rofz}. The solid curve was constructed using the inversion
method and should ideally match the dashed curve. However, the inversion method breaks down for $z\sim0.6$ because there
$ER/M\rightarrow0$, in which case the $dr/dz$ equation becomes
unstable as explained in Subsection~\ref{sub:FRW-example}. To see this
explicitly, we plot $ER/M$ as a function of redshift $z$ in
Fig.~\ref{cap:-ER/M-}.

\subsection{Cosmological Constant without Cosmological Constant}
Until now, we have given two examples of using the inversion method to
reproduce known LTB models. In this subsection, we construct an LTB
model which reproduces an identical luminosity distance to that
produced by an FRW universe with $\Omega_{M}=0.3$ and
$\Omega_{\Lambda}=0.7$ up to a finite redshift.  In particular, we
will set $R(z)=D_{L}(z)/(1+z)^{2}$, with $D_{L}(z)$ given by
Eq.~(\ref{eq:hogfrwlumdist}).

Although the inversion method is unstable for $z\gtrsim z_{c}$, where
$z_{c}$ is a cutoff redshift (whose generic existence is suggested by
the two examples previously presented), we can use the inversion
method for $z<z_{c}$ and then smoothly patch that model ({\it i.e.}
the function $M(r)$) on to a function $M(r)\sim r^{3}.$ Since the
luminosity distance function is sensitive to the cosmological constant
only for $z\lesssim 1$, such models tend to give good fits to the
observed luminosity distance data.

\begin{figure}
\begin{center}\includegraphics{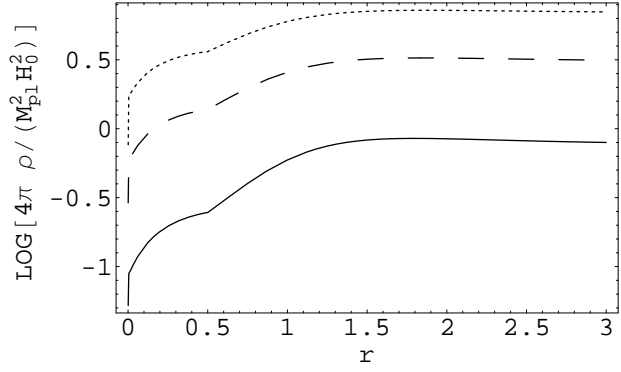}\end{center}

\caption{\label{rhoplot}Using the inversion method, we have
computed the energy density $\rho(t,r)$ corresponding to the
luminosity distance of an FRW universe with $\Omega_{M}=0.3$ and $\Omega_{\Lambda}=0.7$.  The solid, dashed, and dotted curves
correspond to $\rho(t,r)$ at times $t_{\mbox{today}}$,
$t_{\mbox{today}}/2$, and $t_{\mbox{today}}/3$, respectively.  With
$H_0=70$ km/s/Mpc, $t_{\mbox{today}}$ is 13 billion years.  The
redshift corresponding to $r\approx 0.5/H_0$ is $z\approx 0.4$.  The
energy density at $r=0$ is nonzero: $4 \pi
\rho(t_{\mbox{today}},0)/(M_{pl}^2 H_0^2) \sim 10^{-3}$.}
\end{figure}

For our trial example, in addition to setting $R(z)$ according to
Eq.~(\ref{eq:hogfrwlumdist}) and $R_{0}(r)=0$, we choose\[
E(r)=\frac{1}{2} H_{0}^{2}r^{2} \exp[-2 H_0 r], \] which has a local
spatial curvature only for $r\approx 0$.  The solution
${t(z),r(z),M_1(z)}$ is obtained by solving
Eqs. (\ref{eq:lumindist2pr}), (\ref{eq:lumindist3pr}), and
(\ref{eq:dRdzchainrule}) with the initial conditions to the differential
equations set as outlined near Eq.~(\ref{eq:tlinear}).
Explicitly, we take $t(z_i)=0.93/H_0$, $r(z_i)=0.97 z_i/H_0$,
$M_1(z_i)= 0.028 r^3(z_i)$, and $z_i=10^{-5}$.  As previously
discussed near Eq.~(\ref{eq:tlinear}), we solved linearized
equations to obtain $t(z_i)$ and $r(z_i)$, assuming $M_1(z_i)= 0.028
r^3(z_i)$. (The coefficient $0.28$ in $M_1(z_i)$ is related to the
energy density at $r=0$ through Eq.~(\ref{eq:energydensity}).)  The
differential equation was then solved from $z\approx0$ to
$z=z_{c}\approx0.4$.  From approximately this redshift onward
\footnote{The differential equation does not become unstable until
$z\approx 0.505$, but we stopped it earlier to obtain a smoother fit
to the $M(r)\sim r^3$ ansatz.}, the inversion method is unstable
because of the form $0/0$.  (It is unclear from our analysis whether this
instability is purely numerical or whether this instability can be
removed by a reformulation of the equation and perturbations of the
boundary conditions.  We defer this question to a future work.)  Hence,
we fit a smooth $M(r)$ function starting from close to this
point:\begin{eqnarray*} M(r) & = & \left\{
\begin{array}{ccc} M_{1}^{\textrm{num}}(z^{\textrm{num}}(r)) & &
r\leq0.5/H_0\\ C_{1}(r+C_{2})^{3} & &
r>0.5/H_0\end{array}\right.\end{eqnarray*} where the superscript
{}``num'' refers to the numerical solutions to
Eqs. (\ref{eq:lumindist2pr}), (\ref{eq:lumindist3pr}), and
(\ref{eq:dRdzchainrule}) and $C_{1}$ and $C_{2}$ are adjusted to match
the value and the $r$ derivative at $r=0.5$. We then take the
extended $M(r)$ and solve Eq. (\ref{eq:maineqeinst}) to obtain
$R(t,r)$ numerically.  Finally, we then solve
Eqs. (\ref{eq:lumindist1}) and (\ref{eq:lumindist2}) to obtain the
full luminosity distance beyond $z=z_{c}\approx0.4$.  Note that a
functional choice of $C_{1}(r+C_{2})^{3}$ was made to obtain an
approximately homogeneous $\rho(t,r)$ for $r\sim 1/H_0$.

\begin{figure}
\begin{center}\includegraphics{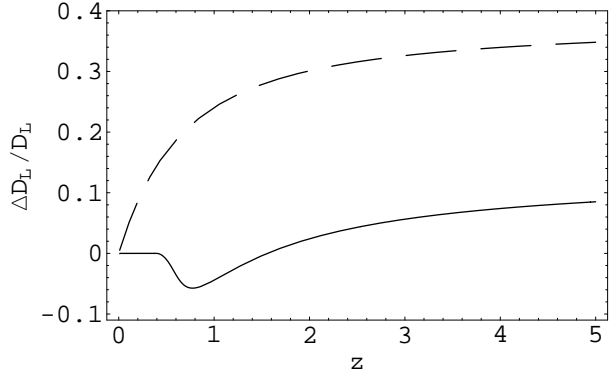}\end{center}

\caption{\label{mainnewmodel}The solid curve corresponds to
$(D_{L}^{\Lambda=0.7}(z)-D_{L}^{\textrm{model}}(z))/D_{L}^{\Lambda=0.7}(z)$
which corresponds to the deviation of the numerically constructed
{}``model'' luminosity distance curve from the FRW curve (where the
FRW model is one with $\{\Omega_{\Lambda}=0.7,\Omega_{M}=0.3\}$).  For
comparison, we plot the dashed curve, which corresponds to
$(D_{L}^{\Lambda=0.7}(z)-D_{L}^{\Lambda=0}(z))/D_{L}^{\Lambda=0.7}(z)$,
where $D_{L}^{\Lambda=0}$ corresponds to a flat matter-dominated FRW
model without any cosmological constant. It is clear that the model
reproduces the luminosity curve exactly from $z=0$ up to $z=0.4$ (by
construction) and there is a less than about 5\% deviation from the
FRW $\{\Omega_{\Lambda}=0.7,\Omega_{M}=0.3\}$ luminosity distance
curve until about $z=3$. Furthermore, the deviation error is seen to
plateau at large redshifts. }
\end{figure}
The resulting solutions can be seen in Figs.~\ref{rhoplot} and
\ref{mainnewmodel}. Clearly, the inhomogeneous model, whose energy
density is shown in Fig~\ref{rhoplot}, reproduces the luminosity
distance of the $\{\Omega_{\Lambda}=0.7,\Omega_{M}=0.3\}$ FRW model
exactly from $z=0$ to $z=0.4$ and the luminosity distance curve
deviation from that of the FRW model is less than around 5\% until
$z=3$. (The $\{\Omega_{\Lambda}=0.7,\Omega_{M}=0.3\}$ FRW model is
known to give a good fit to the supernova data.)

Hence, we have explicitly demonstrated that the LTB model can be tuned
to obtain a luminosity-distance-redshift relationship which accurately
reproduces that of a standard flat FRW universe with a cosmological
constant and dark matter.  This solution differs from previously
proposed solutions in that the luminosity distance curve is exactly
that of the FRW universe with a cosmological constant from $z=0$ to
$z=0.4$.  It is interesting to note that the energy density is
approximately homogeneous for large $r$ but has a void close to $r=0$
(see Fig.~\ref{rhoplot}), similarly to the model of
\cite{Alnes:2005rw}.

\section{\label{sec:Naked-Singularity}Naked Singularity Formation}

In this section, we demonstrate that the LTB model is susceptible to
the formation of naked singularities when $E(r)$ switches sign from
positive to negative as $r$ increases while $M'(r)$ is positive in
that region. More specifically, we consider situations in which
$k(r)\sim-2E(r)/r^{2}$ ({\it i.e.}, the ``local'' spatial curvature
factor) makes a smooth transition from an {}``underdense'' universe to
an {}``overdense'' universe as $r$ crosses $r_{0}$ from below. This
naked singularity may be interpreted as due to formation of caustics
arising from matter accretion.  Since any realistic system has nonzero
pressure at sufficiently small length scales, this singularity is
unphysical and is an artifact of pressureless dust approximation. This
type of naked singularity would develop if one naively smooths out
inhomogeneity profiles such as the one used by \cite{Nambu:2005zn}.

Now, note that because of the expression for the energy density in
Eq. (\ref{eq:energydensity}), for the energy density to be positive,
$M'(r)$ must have the same sign as $R,_{r}$. Using
Eq. (\ref{eq:intermedexp}), which is 
valid for $ER/M\ll1$ (which is almost always true in the vicinity
of $r=r_{0}$, the point at which $E(r)$ changes sign smoothly),
let us consider the situation in which $R_{0}(r)=0$ and $t_{0}=0$
(the case considered by \cite{Nambu:2005zn}). We find\begin{eqnarray}
R,_{r} & = & P(t,r)\left[E'(r)-\frac{E(r)}{3}\frac{M'(r)}{M(r)}+\frac{5\cdot2^{1/3}}{3\cdot3^{2/3}}\frac{M'(r)}{t^{2/3}M^{1/3}(r)}\right]\label{eq:rder}\end{eqnarray}
where\begin{equation}
P(t,r)=\frac{3\cdot3^{1/3}}{5\cdot2^{2/3}}\frac{t^{4/3}}{M^{1/3}(r)}.\end{equation}
Supposing that the last term proportional to $M'(r)/(t^{2/3}M^{1/3})$ can
be neglected compared to the $E'(r)$ term and $M'(r)>0$, we have
the condition\begin{equation}
E'(r)>\frac{E(r)}{3}\frac{M'(r)}{M(r)}\end{equation}
if $M>0$. If we choose a smooth $E(r)$ which goes from a
positive value to a negative value as $r$ increases past $r_{0}$, 
the Taylor expansion of $E(r)$ is then required to have the form \begin{equation}
E(r)=E_{1}(r_{0})(r-r_{0})+\frac{1}{2}E_{2}(r_{0})(r-r_{0})^{2}+...\end{equation}
 with $E_{1}<0$, where $E_{n}$ corresponds to the $n$th derivative
of $E(r)$. Hence, in a sufficiently small neighborhood of $r=r_{0}$,
for $E_{1}(r_{0})\neq0$ we have the condition\begin{equation}
\frac{(r-r_{0})}{3}\frac{M'(r)}{M(r)}>1.\end{equation}
 If we assume $M'/M$ does not switch sign at $r=r_{0}$, this condition
is clearly not satisfied in the vicinity of $r_{0}$. This means that when
the $E'(r)$ term dominates the $R,_{r}$ expression, the energy density cannot
remain positive definite.

Now, suppose that the $M'(r)/(t^{2/3}M^{1/3})$ term dominates over the $E'$
term in Eq. (\ref{eq:rder}), which would certainly be true near the
big-bang singularity at $t\rightarrow0$. We find the following condition for $M'>0$
and $M>0$\begin{equation}
\frac{5\cdot2^{1/3}}{3^{2/3}}\frac{M^{2/3}(r)}{t^{2/3}}>E(r),\end{equation}
which can be satisfied in the vicinity of $r=r_{0}$.  This means that
energy can be positive definite in these regime. Hence, we arrive
at a naively puzzling question why the energy density, which is initially
positive definite near the big bang singularity, develops into a negative
energy density when $E'(r)$ term governs the value of $\rho$. The
answer is that a naked singularity develops during the course of matter
evolution.

To see this in a more obvious way, let us take a concrete model of\begin{equation}
M(r)=\rho_{0}\frac{a_{0}^{3}r^{3}}{6}\label{eq:Mstep}\end{equation}
 and\begin{equation}
E(r)=\frac{-r^{2}}{2L^{2}}\tanh(\kappa\frac{(r-r_{0})}{L}),\label{eq:Estep}\end{equation}
which represents a version of the model of \cite{Nambu:2005zn} with
the step function smoothed out. We can explicitly solve for the
coordinate where $R,_{r}$
becomes negative by solving $R,_{r}=0$ :\begin{equation}
\frac{10}{3\cdot6^{1/3}}\frac{L^{2}a_{0}^{2}\rho_{0}^{2/3}}{t^{2/3}}-\frac{\kappa r}{L}\textrm{sech}^{2}[\frac{\kappa(r-r_{0})}{L}]-\tanh[\frac{\kappa(r-r_{0})}{L}]=0.\end{equation}
Expanding about $r=r_{0}$ to quadratic order in $r-r_{0}$, we find
the radius at which $R,_{r}$ becomes zero to be\begin{eqnarray}
\frac{r_{s}-r_{0}}{L} & = & \frac{L}{\kappa^{2}r_{0}}\pm\frac{1}{\kappa}\sqrt{\frac{-5\cdot6^{2/3}}{9\kappa}\frac{L}{r_{0}}[a_{0}^{2}L^{2}]\rho_{0}^{2/3}t^{-2/3}+\frac{L^{2}}{\kappa^{2}r_{0}^{2}}+1}.\end{eqnarray}
Requiring that the solution be real results in the condition\begin{equation}
t\gtrsim\sqrt{5}[\rho_{0}a_{0}^{3}L^{3}](\frac{L}{\kappa r_{0}})^{3/2}\label{eq:zerodevelopeq},\end{equation}
after which time, $R,_{r}$ has a value of zero near $r\approx r_{0}+\frac{L^{2}}{\kappa^{2}r_{0}}$.
Comparing this time with the time $t_c$ at which the overdense part of the
universe starts to collapse ($t_{c}\approx\frac{\pi
  a_{0}^{3}L^{3}\rho_{0}}{6})$, we see that if $\kappa\gg1$ while $\mathcal{O}(r_{0})\sim\mathcal{O}(L)$, $R,_{r}$ will reach zero before $t$ reaches $t_{c}$. 

\begin{figure}
\begin{center}\includegraphics{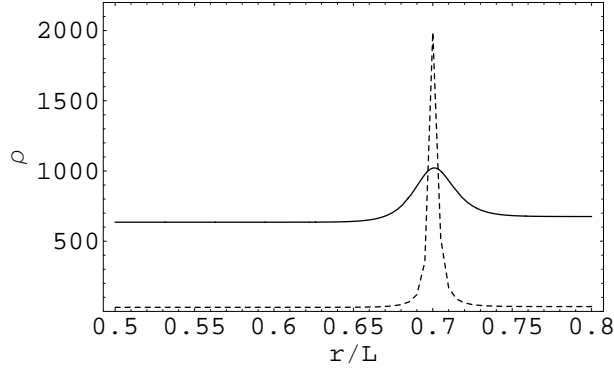}\end{center}

\caption{\label{fig:divergenceofenergy} The energy density is plotted as
a function of radius for times $t=0.01L$ (solid) and $t=0.042L$
(dotted) for the model of Eqs.~(\ref{eq:Mstep}) and (\ref{eq:Estep})
with $\{ r_{0}=0.7L,\kappa=50,\rho_{0}=3.5\frac{1}{L^{2}}\}$. As
expected, the energy density diverges as $R,_{r}\rightarrow0$ at
$r\approx r_{0}$ near the time $t\gtrsim0.04L$. Note also that the energy
density for $r>r_{0}$ is larger than $r<r_{0}$, since one side is
overdense while the other side is underdense.}
\end{figure}

When $R,_{r}$ reaches zero at nonzero $r=r_{s}$ without $M'(r)$
vanishing, we have a Ricci curvature singularity there. We have also
checked the divergence of the energy density near the spacetime region
of the naked singularity for several numerical examples, one of which
is shown in Fig.~\ref{fig:divergenceofenergy}. Since no metric element
became zero before the singularity comes into existence, we see that
the curvature singularity is not protected by a horizon; {\it i.e.}, it
is a naked singularity. As is well known \cite{waldbook}, naked singularities
can develop with perfect fluid systems. Hence, if $E(r)$ changes
sign at $r=r_{0}$ with $M(r)>0$ and $M'(r)>0$ in the vicinity of
$r_{0}$, a naked singularity develops, and the Einstein equations break
down in that region. Furthermore, if
we naively continue using the Einstein equations past the development
of naked singularities, one finds a negative energy density region
which is unphysical as expected.

Here we note that the model of \cite{Alnes:2005rw} does not have any
naked singularity problems associated with shell crossing. To see
this, we will use the analytic approximation of
Eq.~(\ref{eq:intermedexp}) applicable to the region in which
$|E(r)R/M|\ll1$. In the scenario specified by Eqs.~(\ref{eq:params}),
we have \begin{equation}
|\frac{E(r_{0})R(t,r_{0})}{M(r_{0})}|\approx|0.3(10^{-4}+4H_{0}t)^{2/3}(1-\frac{2\times10^{-8}}{10^{-4}+4H_{0}t}+0.06[10^{-4}+4H_{0}t]^{2/3})|\ll1\end{equation}
until the time $t>1/H_{0}$. One can then use
Eq. (\ref{eq:intermedexp}) in this regime to evaluate $R,_{r}/r$ and
find that the energy density does not diverge anywhere. 
The key reason is that $E(r)$ never switches sign, unlike the previous case.

\begin{figure}
\begin{center}\includegraphics{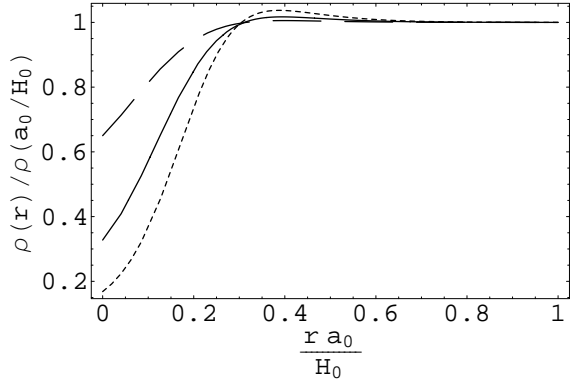}\end{center}

\caption{\label{cap:Energy-density-as}The energy density as a function of radius
for three different times. The dotted curve corresponds to $t=0$ (recombination),
the dashed curve corresponds to $t=0.1/H_{0}$, and the solid curve corresponds
to $t=0.5/H_{0}.$ }
\end{figure}

There is a peculiar physical feature of the energy density as a function
of time, as shown in Figure~\ref{cap:Energy-density-as}.%
Some of the matter density from the
transition region near $r_{0}$ initially transfers to the $r=0$
region before being diluted away by curvature dominated expansion.
In other words, the density near $r=0$ is not a monotonic function
of time because the initial conditions are set up such that the fluid
is flowing towards $r=0$.

\section{Conclusion}

In this paper, we have derived a set of differential equations in the
context of LTB cosmologies (spherically symmetric pressureless dust
solutions to the Einstein equations) which can be numerically solved
to obtain almost any luminosity-distance-redshift relationship that
can be produced by a homogeneous and isotropic FRW model.  We have
solved this set of differential equations for several examples to
demonstrate the feasibility of our inversion method.  Unlike many
other methods in the literature, our method can be used to dial in the
geometry that generates the desired luminosity distance {\em exactly} in a
finite redshift interval.  We have also given explicit examples of
naked singularity formation in LTB cosmologies in the region where
$E(r)$ changes sign.
 
Although our work serves as a step towards building cosmological
models competitive to the standard FRW cosmology, the research program
is far from completion. Toy models such as the LTB cosmologies are
arguably not yet convincing contenders for compelling alternatives to
standard inflationary cosmology, since being at the
center of the universe requires giving up the Copernican principle
(though recently there has been some effort to alleviate this problem
\cite{Kai:2006ws}) and there is not yet a convincing structure
formation history that could explain such radial inhomogeneities.
Nonetheless, it is not obvious whether this class of LTB models can be
ruled out from current observations \cite{Alnes:2005rw}.  Furthermore,
there is some evidence that our galaxy is in a void, qualitatively
similar to the voids presented in two of our numerical examples.
However, the existence of the void is still currently being
investigated (see for example \cite{Frith:2005et}), and whether the
magnitude of the void is sufficient to cause the redshift effects
presented in this paper is unclear.  Regarding tests of LTB
cosmologies, any physical probe testing the isotropy of the universe
from a point separated sufficiently far away from us could in
principle be useful.  We leave this problem for future investigation.

\begin{acknowledgments}
We thank E. Kolb for discussions and suggesting this problem to
us. DJHC also thanks D. McCammon for useful discussions and
L. Everett for comments on the manuscript.  The work of DJHC and AER
was supported in part by the DOE Outstanding Junior Investigator Program
through Grant No. DE-FG02-95ER40896 and NSF Grant No. PHY-0506002. AER
was supported in part by the Wisconsin Alumni Research Foundation.
\end{acknowledgments}

\begin{appendix}
\section{FRW Luminosity distance}

The FRW luminosity distance in flat FRW universe with pressureless
dust fraction $\Omega_{M}$ and cosmological constant fraction
$\Omega_{\Lambda}$ without spatial curvature
($\Omega_{M}+\Omega_{\Lambda}=1$) is given by\begin{equation}
D_{L}(z)=\frac{1+z}{H_{0}}\int_{1}^{1+z}dyI(y)\label{eq:hogfrwlumdist}\end{equation}
\begin{equation}
I(y)\equiv\frac{1}{\sqrt{\Omega_{\Lambda}+(1-\Omega_{\Lambda})y^{3}}}\end{equation}
where the integral is numerically a value of order $1$ with a logarithmic
dependence on $z$. Note that a Taylor expansion is not very efficient
approximation of the integral as fifth order expansion gives a good
fit to only about $z\approx1.5$.

To check the plausibility of this expression, consider the solvable
case of matter dominated universe $\Omega_{\Lambda}=0$:\begin{eqnarray}
D_{L}(z) & = & \frac{1+z}{H_{0}}\int_{1}^{1+z}dyy^{-3/2}\\
 & = & 2\frac{1+z}{H_{0}}[1-\frac{1}{\sqrt{1+z}}]\end{eqnarray}
which is the familiar expression. If this is matched to LTB model,
we would write\begin{eqnarray}
R(z) & = & \frac{D_{L}(z)}{(1+z)^{2}}\\
 & = & 2\frac{1}{H_{0}}[\frac{1}{1+z}-\frac{1}{(1+z)^{3/2}}].\end{eqnarray}

\section{Explicit Solution}

Einstein equation (\ref{eq:einstein00orig}) can be solved by constructing
characteristic curves. The result can be expressed as \begin{equation}
-(t-t_{0})\sqrt{2}E(r)+\sqrt{R(t,r)}\sqrt{E(r)R(t,r)+M(r)}-Q(r)=\frac{M(r)}{\sqrt{E(r)}}\ln[\frac{\sqrt{R(t,r)}+\sqrt{\frac{M(r)}{E(r)}+R(t,r)}}{\sqrt{R_{0}(r)}+\sqrt{\frac{M(r)}{E(r)}+R_{0}(r)}}]\label{eq:maineqeinst}\end{equation}
\begin{equation}
Q(r)\equiv\sqrt{R_{0}(r)}\sqrt{E(r)R_{0}(r)+M(r)}\end{equation}
for $E>0$ while for $E<0$, we have\begin{equation}
F(r)-(t-t_{0})\sqrt{2}E(r)=\frac{M(r)}{\sqrt{-E(r)}}\arcsin[\sqrt{\frac{-E(r)}{M(r)}}\sqrt{R(t,r)}]-\sqrt{R(t,r)}\sqrt{E(r)R(t,r)+M(r)}\label{eq:E<0}\end{equation}
\begin{equation}
F(r)=\frac{M(r)}{\sqrt{-E(r)}}\arcsin[\sqrt{\frac{-E(r)}{M(r)}}\sqrt{R_{0}(r)}]-\sqrt{R_{0}(r)}\sqrt{E(r)R_{0}(r)+M(r)}.\end{equation}
Here, $R_{0}(r)$ is the initial condition specification for $R(t,r)$:
i.e. $R(t_{0},r)=R_{0}(r)$. Note that the solution can be written
a little more explicitly in the limit $|EM/R|\ll1$ and $|EM/R|\gg1$.
For $|EM/R|\ll1$, we find\begin{equation}
R=\left(R_{0}^{3/2}+3\sqrt{\frac{M(r)}{2}}(t-t_{0})\right)^{2/3}\left(1+\frac{E(r)}{M(r)}\delta\right)\label{eq:intermedexp}\end{equation}
\begin{equation}
\delta=\frac{2\cdot2^{1/3}R_{0}^{3}+6\cdot2^{5/6}\sqrt{M}R_{0}^{3/2}(t-t_{0})+9\cdot2^{1/3}M(t-t_{0})^{2}}{5[2R_{0}^{3/2}+3\sqrt{2M}(t-t_{0})]^{4/3}}-\frac{2R_{0}^{5/2}}{5[2R_{0}^{3/2}+3\sqrt{2M}(t-t_{0})]}\end{equation}
For $ER/M\gg1$, the leading order self-consistency requires\begin{equation}
R=R_{0}+\sqrt{2E}(t-t_{0})+\frac{M}{2E}\ln[\frac{R}{R_{0}}]+\mathcal{O}(\frac{1}{E^{2}}).\end{equation}
Hence, we can approximate in the regime of interest \begin{equation}
R\approx\sqrt{2E(r)}(t-t_{0})+R_{0}(r)+\frac{M(r)}{2E(r)}\ln\left(1+\frac{\sqrt{2E(r)}}{R_{0}(r)}(t-t_{0})\right).\label{eq:solinunderdense}\end{equation}

\end{appendix}


\begin{thebibliography}{99}

\bibitem{Moffat:2005yx}
  J.~W.~Moffat,
  JCAP {\bf 0510}, 012 (2005)
  [arXiv:astro-ph/0502110].

\bibitem{Nambu:2005zn}
  Y.~Nambu and M.~Tanimoto,
  arXiv:gr-qc/0507057.

\bibitem{Ishibashi:2005sj}
  A.~Ishibashi and R.~M.~Wald,
  Class.\ Quant.\ Grav.\  {\bf 23}, 235 (2006)
  [arXiv:gr-qc/0509108].

\bibitem{Alnes:2005rw}
  H.~Alnes, M.~Amarzguioui and O.~Gron,
  Phys.\ Rev.\ D {\bf 73}, 083519 (2006)
  [arXiv:astro-ph/0512006].

\bibitem{Vanderveld:2006rb}
  R.~A.~Vanderveld, E.~E.~Flanagan and I.~Wasserman,
  Phys.\ Rev.\ D {\bf 74}, 023506 (2006)
  [arXiv:astro-ph/0602476].

\bibitem{Apostolopoulos:2006eg}
  P.~S.~Apostolopoulos, N.~Brouzakis, N.~Tetradis and E.~Tzavara,
  JCAP {\bf 0606}, 009 (2006)
  [arXiv:astro-ph/0603234].

\bibitem{Kai:2006ws}
  T.~Kai, H.~Kozaki, K.~i.~nakao, Y.~Nambu and C.~M.~Yoo,
  arXiv:gr-qc/0605120.

\bibitem{Moffat:2005ii}
  J.~W.~Moffat,
  JCAP {\bf 0605}, 001 (2006)
  [arXiv:astro-ph/0505326].

\bibitem{Celerier:1999hp}
  M.~N.~Celerier,
  Astron.\ Astrophys.\  {\bf 353}, 63 (2000)
  [arXiv:astro-ph/9907206].

\bibitem{Biswas:2006ub}
  T.~Biswas, R.~Mansouri and A.~Notari,
  arXiv:astro-ph/0606703.

\bibitem{Chuang:2005yi}
  C.~H.~Chuang, J.~A.~Gu and W.~Y.~Hwang,
  arXiv:astro-ph/0512651.


\bibitem{Riess:1998cb}
  A.~G.~Riess {\it et al.}  [Supernova Search Team Collaboration],
  Astron.\ J.\  {\bf 116}, 1009 (1998)
  [arXiv:astro-ph/9805201].

\bibitem{Perlmutter:1998np}
  S.~Perlmutter {\it et al.}  [Supernova Cosmology Project Collaboration],
  Astrophys.\ J.\  {\bf 517}, 565 (1999)
  [arXiv:astro-ph/9812133].

\bibitem{Sugiura:1999dp}
  N.~Sugiura, K.~i.~Nakao and T.~Harada,
  Phys.\ Rev.\ D {\bf 60}, 103508 (1999)
  [arXiv:gr-qc/9911090].

\bibitem{Mustapha:1997xb}
  N.~Mustapha, B.~A.~Bassett, C.~Hellaby and G.~F.~R.~Ellis,
  Class.\ Quant.\ Grav.\  {\bf 15}, 2363 (1998)
  [arXiv:gr-qc/9708043].


\bibitem{Kristian:1965sz}
  J.~Kristian and R.~K.~Sachs,
  Astrophys.\ J.\  {\bf 143}, 379 (1966).


\bibitem{waldbook}
R.~M.~Wald, {\it General Relativity} (Univ. Chicago Press, Chicago, IL 1984).

\bibitem{Frith:2005et}
  W.~J.~Frith, N.~Metcalfe and T.~Shanks,
  arXiv:astro-ph/0509875.


\end{thebibliography}
\end{document}